\documentclass[]{llncs}
\usepackage{graphicx}
\usepackage{amsmath,amssymb} 
\usepackage{color}
\usepackage{array}
\usepackage{comment} 
\usepackage{caption}
\usepackage{authblk}

\newcommand\blue[1]{\textcolor{black}{#1}}

\begin{document}


\def\x{{\mathbf x}}
\def\p{{\mathbf p}}
\def\X{{\mathbf X}}
\def\v{{\mathbf v}}
\def\d{{\mathbf d}}
\def\xh{{\hat{\x}}}
\def\uh{ \bar{\mathbf{u}}}
\def\u{ \mathbf{u}}
\def\r{ \mathbf{r}}
\def\y{{\mathbf y}}
\def\Y{{\mathbf Y}}
\def\z{{\mathbf z}}
\def\Z{{\mathbf Z}}
\def\J{{\mathbf J}}
\def\u{{\mathbf u}}
\def\U{{\mathbf U}}
\def\Q{{\mathbf Q}}
\def\I{{\mathbf I}}

\def\G{{\mathbf G}}
\def\D{{\mathbf D}}
\def\Dh{{\hat{\D}}}
\def\g{{\mathbf g}}
\def\A{{\mathbf A}}
\def\B{{\mathbf B}}
\def\W{{\mathbf W}}
\def\w{{\mathbf w}}

\def\M{{\mathcal{M}}}
\def\P{{\mathcal{P}}}
\def\ReLU{{\text{ReLU}}}
\def\b{{\mathbf b}}
\def\S{{\mathcal{S}}}
\def\H{{\mathcal{H}}}
\def\C{{\mathbf C}}
\def\L{{\mathbf L}}
\def\Ai{\A_i}
\def\T{{\mathrm T}}
\def\Ti{\T_i}
\def\R{{\mathbf R}}
\def\PP{{\mathbf P}}
\def\RR{{\mathbb R}}
\def\SS{{\mathbf S}}
\def\V{{\mathbf V}}
\def\c{{\mathrm c}}
\def\AA{{\mathbf A}}
\def\BB{{\mathbf B}}
\def\FF{{\mathbf F}}
\def\bb{{\mathbf b}}
\def\E{{\mathbf E}}


\title
{The Rate-Distortion-Accuracy Tradeoff: \\ JPEG Case Study}

\institute{}

\makeatletter
\newcommand{\printfnsymbol}[1]{%
  \textsuperscript{\@fnsymbol{#1}}%
}
\makeatother

\author[]{Xiyang Luo\thanks{Equal contribution.}}
\author[]{Hossein Talebi\printfnsymbol{1}}
\author[]{Feng Yang}
\author[]{Michael Elad}
\author[]{Peyman Milanfar}
\affil[]{Google Research}
\date{}                     

\maketitle


\vspace{-0.2in}
\begin{figure}[]
\begin{center}
    \begin{tabular}{c c  c c}
    \multicolumn{2}{c}{Rate-Distortion Optimization} & \multicolumn{2}{c}{Rate-Accuracy Optimization} \\ 
    {\scriptsize Default} & {\scriptsize Optimized} & {\scriptsize	Default}& {\scriptsize	Optimized}  \\
    \includegraphics[ width=0.20\linewidth]{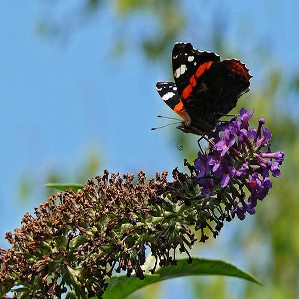} \vspace {5pt} &
    \includegraphics[ width=0.20\linewidth]{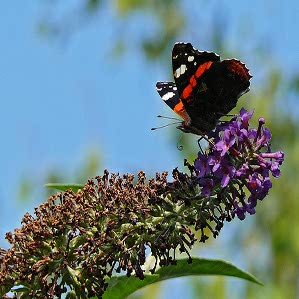} & 
    \includegraphics[ width=0.20\linewidth]{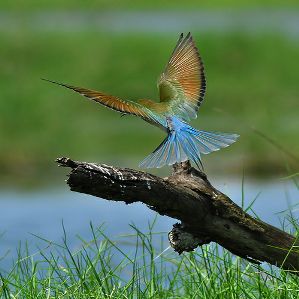}&
    \includegraphics[ width=0.20\linewidth]{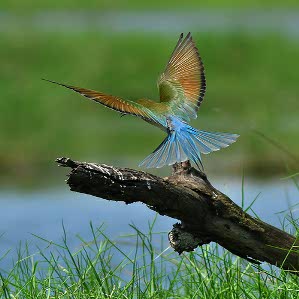} \\
    {\scriptsize PSNR = 30.24dB} & {\scriptsize	PSNR = 30.62dB} & {\scriptsize ``Dragonfly" [0.16]} & {\scriptsize ``Bee Eater" [0.25]} \\ BPP = 3.06 & BPP = 2.30  & BPP = 1.706 & BPP = 1.680 
    \end{tabular}
    \end{center}
    \caption{We offer an optimization of JPEG's quantization tables for improved rate-distortion or rate-accuracy performance. The above are sample results showing (i) a $17\%$ reduction in file size while maintaining the same quality, and (ii), a corrected classification while saving $1.5\%$ in file size. }\label{fig:teaser}
\end{figure}

\vspace{-0.35in}
\begin{abstract}
Handling digital images is almost always accompanied by a lossy compression in order to facilitate efficient transmission and storage. This introduces an unavoidable tension between the allocated bit-budget (rate) and the faithfulness of the resulting image to the original one (distortion). An additional complicating consideration is the effect of the compression on recognition performance by given classifiers (accuracy). This work aims to explore this rate-distortion-accuracy tradeoff. As a case study, we focus on the design of the quantization tables in the JPEG compression standard. We offer a novel optimal tuning of these tables via continuous optimization, leveraging a differential implementation of both the JPEG encoder-decoder and an entropy estimator. This enables us to offer a unified framework that considers the interplay between rate, distortion and classification accuracy. In all these fronts, we report a substantial boost in performance by a simple and easily implemented modification of these tables.



\end{abstract}


\section{Introduction}

Digital images are almost always compressed, exploiting their massive spatial and statistical redundancies in order to save storage space and/or transmission rates. The common practice is to use standard \emph{lossy} coding formats, such as JPEG, JPEG-2000, HEIF,  
or others. Lossy compression implies a permitted deviation between the resulting compressed-decompressed image and its original version. This error can be controlled by the bit-budget given to the image, creating the well-known rate-distortion tradeoff, which is at the very foundation of information theory~\cite{cover2012elements}. 

If these images are to be fed to a classification machine for recognition purposes, the compression distortion may induce errors in the decisions made.
In such scenarios we are to consider three performance measures that are at odds with each other: rate, distortion, and classification accuracy. This work focuses on this rate-distortion-accuracy, tradeoff, aiming to show that improved compression performance are within reach while preserving the standard coding paradigm. 

As a case study, our paper focuses on
JPEG compression. Among the various available image coding methods, JPEG holds a unique status, being the most commonly used and widely spread. This image format\footnote{JPEG is a \emph{decompression} format, leaving some freedom in the design of the encoder.} is the de-facto default in digital cameras and cell-phones, in all browsers, and in every image editing software package. JPEG's popularity could be attributed to its relative simplicity, hardware friendliness, reasonable rate-distortion performance, and beyond all these, the perfect timing it had in getting to the market. And so, while much better-performing compression algorithms are already available, JPEG's dominance of the market does not seem to be challenged in the near future. 

\begin{figure}[t!]
\vspace{-0 mm}
\begin{center}
\includegraphics*[viewport=1 220 1000 400,scale=0.48]{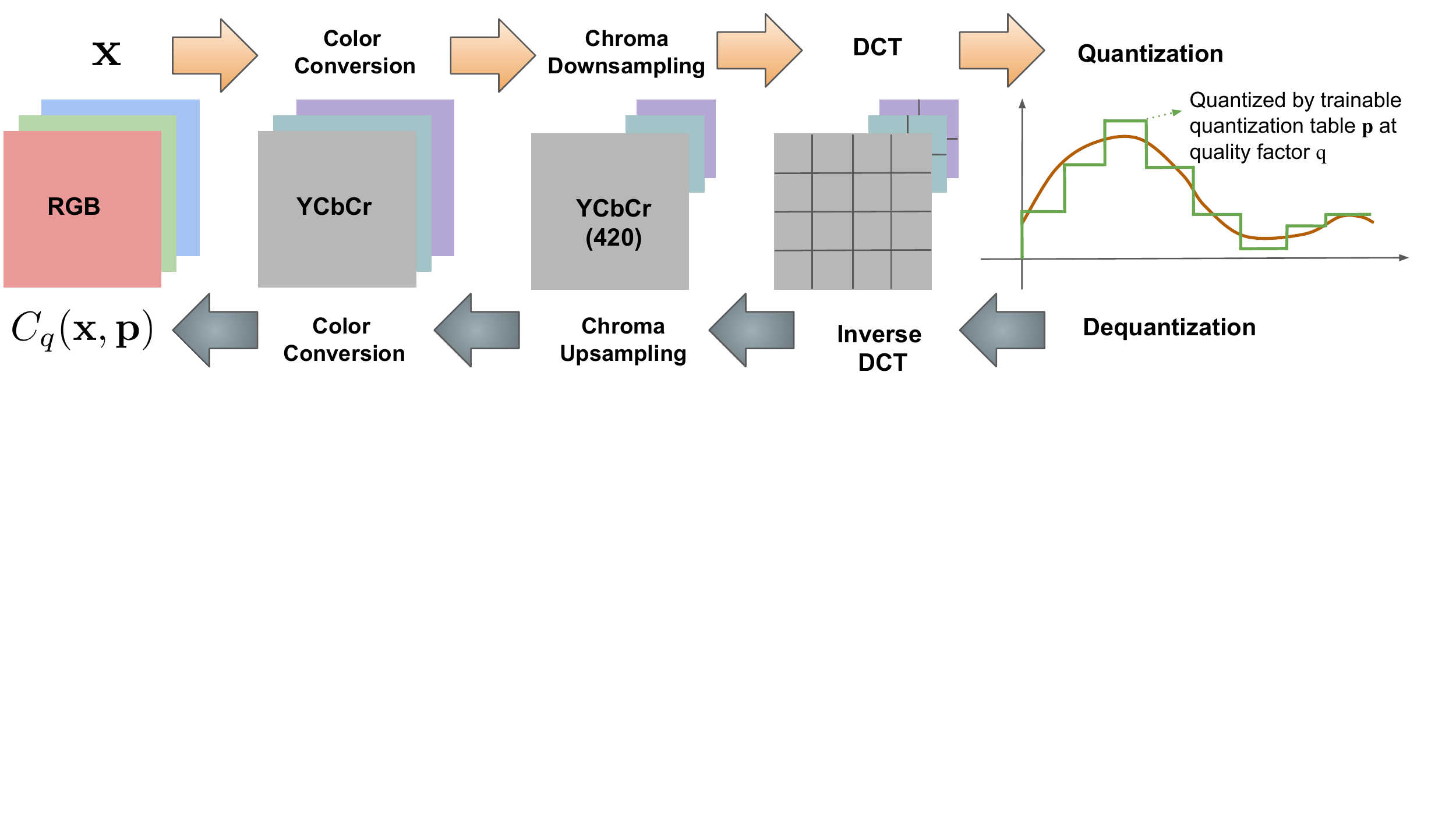}
\end{center}
\vspace{-6 mm}
{\caption{\small JPEG for an image $\textbf{x}$: In the compression pass, after color conversion, an optional chroma downsampling is applied. Then, DCT is computed on  $8\times8$ blocks for each color channel. The DCT coefficients are quantized using the table $\textbf{p}$ and a scalar factor $q$. Going back through each of these steps results in the decompressed image $C_q(\textbf{x}, \textbf{p})$. Our framework relies on the differentiability of all the JPEG steps in order to enable continuous optimization of the quantization tables.} \label{fig:jpeg_framework}}
\vspace{-2 mm}
\end{figure}

This popularity has motivated past and present attempts to extract the best performance from JPEG while preserving its essence. In this work we target the choice of the two quantization tables used within the JPEG coding process (see Figure~\ref{fig:jpeg_framework}). The Luma and the Chroma are quantized in the DCT domain while operating on $8\times 8$ blocks. The relative quantization step-sizes for each coefficient are stored in these two $8\times 8$ tables. Most JPEG packages offer default values, and many vendors adopt these as is. Are these default tables the best possible ones? As we show in this paper, the answer is negative, and room exists for an improvement of JPEG by redesigning these tables in various ways.  

Previous work from the early 90's and since have already identified the potential in better designing the JPEG quantization tables, considering various approaches~\cite{farid2006digital,chang2002steganographic,monro1993optimum,wu1993rate,ratnakar1995rd,fung1995design,fong1997designing,costa2005identification,yang2008joint,zhang2016just,tuba2017jpeg,hopkins2017simulated,watson1993dct,safranek1994jpeg,shohdohji1999optimization,wang2001designing,jiang2011jpeg,jeong2006jpeg,konrad2009evolutionary,duan2012optimizing,makar2012gradient,chao2013design}. The main effort has been directed to rate-distortion performance improvement, using derivative-free optimization techniques. More recent work also considered tuning these tables for better recognition results. More on these is  described in Section \ref{Sec2:PreviousWork}. While addressing the same general goals, the approach we take in this paper is markedly different. We offer a continuous optimization strategy for tuning these two tables, while considering the above-mentioned three performance measures: rate, distortion and accuracy. 

Our work considers two different design goals and two modes of optimization. As for the design goals, we consider both rate-distortion and rate-accuracy objectives -- the first aims to set the quantization tables for getting the smallest $L_2$-error after compression-decompression for any given bit rate, while the second sets those tables so as to provide the most accurate recognition-rates. We address these goals by considering two optimization setups: universal and per-image modes of work. In the universal case we optimize the choice of the quantization tables for a large corpus of images, essentially proposing a replacement to the commonly-used default values. The second setup aims to fit the best tables for each image so as to extract better JPEG performance. 

Broadly speaking, we formulate each of the above design problems as a non-convex yet smooth optimization task, where the loss to be minimized varies from one case to another. In all cases, the JPEG encoding, decoding and its bit-rate evaluation are all replaced with differentiable proxies. In the classification case, the loss includes a penalty for the accuracy of ResNet~\cite{he2016identity} (or MobileNet~\cite{howard2019searching}) over the ImageNet dataset. The optimization itself 
is performed using a mini-batch gradient descent algorithm,  
and a use of back-propagation. 

Extensive experiments presented in this paper expose the surprising ability to substantially improve JPEG performance in the two considered scenarios. See Figure~\ref{fig:teaser} for two illustrative examples. In terms of rate-distortion, we show a gain of up to 25\% in file-size while maintaining the same image quality (measured in PSNR). Similarly, we show an ability for an increase of up to $0.7\%$ in classification accuracy. \blue{Our experiments show that the optimized tables for MobileNet are just as effective for ResNet, implying that one optimized set may serve various recognition/classification architectures.} We note that our overall methodology could easily be fitted to other compression standards by constructing their differentiable implementation and defining their parameters to be tuned. 



\section{Related Work}
\label{Sec2:PreviousWork}

The important role that the quantization tables play in JPEG has been recognized and exploited in past work for forensics, steganography and more (e.g. \cite{farid2006digital,chang2002steganographic}). In this paper we focus on improving JPEG performance by re-tuning these tables, a topic that has been investigated in past work as well. In the following we briefly account for the relevant literature on this subject, emphasizing the objectives targeted and the means (i.e. algorithms) for getting their results. 

An obvious and expected line of work has dealt with a direct attempt to improve JPEG rate-distortion performance~\cite{monro1993optimum,wu1993rate,ratnakar1995rd,fung1995design,fong1997designing,costa2005identification,yang2008joint,zhang2016just,tuba2017jpeg,hopkins2017simulated}. Papers offering such a treatment differ mainly in the optimization strategy adopted, as the techniques used include simulated annealing~\cite{monro1993optimum,hopkins2017simulated}, coordinate-descent~\cite{wu1993rate,fung1995design,fong1997designing,yang2008joint},  dynamic programming~\cite{ratnakar1995rd}, genetic and evolutionary algorithms~\cite{costa2005identification}, exhaustive separable search~\cite{zhang2016just} and a swarm intelligence method~\cite{tuba2017jpeg}.  Note that all these methods employ derivative-free optimization strategies due to the complex end-to-end function being treated. The work reported in \cite{fung1995design} stands out in this group, as it uses the coordinate-descent approach for targeting an image-adaptive adjustment of the quantization tables.

A related line of activity tunes the quantization tables for better visual quality or improved matching to the human visual system~ 
\cite{watson1993dct,safranek1994jpeg,shohdohji1999optimization,wang2001designing,jiang2011jpeg}. The core idea behind these papers is to optimize the tables while observing the output quality, assessed either via a simplified model of the human visual system, or by relying on subjective tests.

A recent group of papers has been looking at ways to adjust JPEG such that recognition tasks are better served~\cite{jeong2006jpeg,konrad2009evolutionary,duan2012optimizing,makar2012gradient,chao2013design,li2020optimizing}. These papers span a range of decision tasks and optimization techniques. \cite{jeong2006jpeg} uses a direct rate-distortion optimization on a dedicated face image dataset in order to better handle face recognition. Both \cite{konrad2009evolutionary} and \cite{duan2012optimizing} use an evolutionary algorithm, the first for better recognition of eye iris 
images, and the second optimized for visual search results via pairwise image matching. \cite{makar2012gradient,chao2013design} consider general scale-space feature detection accuracy, and optimize the quantization tables using simple frequency domain considerations. In this context, we also mention a parallel body of work that touches on the same goal of improving classification results, referring to alternative compression methods~\cite{liu2018deepn,chamain2019quannet,chamain2019faster}.

Our work differs from the above in two distinct ways. First, as we replace JPEG with a differentiable proxy, we can use continuous optimization methods. Adopting a deep-learning point of view, we use the mini-batch gradient descent and back-propagation, which provide a better potential to reach deeper minima values. Second, our treatment is general, fusing the above and more modes of design into one holistic scheme. Indeed, our work could be considered as an extension of the broad view in \cite{liu2019transferable,suzuki2019image,talebi2020better} that proposed an optimization of a general image pre-processing stage while using a recognition-related or other losses. 


\section{The Proposed Methodology}
\label{Sec3:Proposed}

We now describe our methodology for the optimal design of the quantization tables. We start by introducing our notations.

\subsection{Differentiable JPEG}

 We denote by $C_q(\x,\p):[\mathbb{R}^N \times \mathbb{R}^n] \rightarrow \mathbb{R}^N$ the JPEG compression-decompression of the image $\x \in\mathbb{R}^N$ using quality factor $q$ and quantization tables $\p\in\mathbb{R}^n$. The compression-decompression process is illustrated in Figure~\ref{fig:jpeg_framework}. $C_q(\x,\p)$ is used within our loss function and thus it should be differentiable. Our implementation\footnote{Our implementation refers to the \textsf{YUV420} and \textsf{YUV444} Luma-Chroma sub-sampling, but it can easily be adapted to alternative options.} follows the one reported in \cite{shin2017jpeg}. Next we explain each step of the differentiable JPEG encoder/decoder shown in Figure~\ref{fig:jpeg_framework}.
 
\begin{enumerate}
    \item{Color conversion: An RGB image is converted YUV color space. Since this color conversion is a matrix multiplication, its derivatives are well defined.}
    \item{Chroma downsampling/upsampling: The YUV image can represent full chroma (YUV444), or subsampled CHROMA values (YUV420). The downsampling operation is a $2\times2$ average pooling (YUV444 to YUV420), and the upsampling is implemented with Bilinear interpolation (YUV420 to YUV444).}
    \item{DCT and inverse DCT: The DCT coefficients ($d_{i}$) are computed for $8\times8$ image blocks of each YUV color channel, separately. Note that the DCT operation and its inverse are matrix multiplications, and hence differentiable.}
    \item{Quantization/Dequantization: The DCT quantization $\lfloor \frac{d_{i}}{p_{i}} \rceil$ is performed by $8\times8$ tables $\p$. Note that rounding operation $\lfloor.\rceil$ has zero derivative almost everywhere, and consequently can not be used in our gradient-based learning framework. To alleviate this problem, as Shin et al \cite{shin2017jpeg} suggested, a third-order polynomial approximation of the rounding operation as $\lfloor \frac{d_{i}}{p_{i}} \rceil + (\frac{d_{i}}{p_{i}}-\lfloor \frac{d_{i}}{p_{i}} \rceil)^3$ can be used.}
\end{enumerate}

\subsection{Entropy Prediction}

 We define the function $B_q(\x,\p): \mathbb{R}^N \rightarrow \mathbb{R}^+$ that returns an estimate of the bit-rate consumed for JPEG compressing of $\x $ using quality factor $q$ and quantization tables $\p$. Recall that when using JPEG with a fixed quality factor, the bit-rate is unknown as it depends on the input image in a non-trivial way. For the function $B_q(\cdot)$, we use the entropy estimator proposed in \cite{balle2018variational}, which operates on the quantized DCT coefficients.
    
The approximated entropy can be expressed as
\begin{eqnarray}\label{eq:entropy1}
 E(\widehat{\textbf{d}}) = -\mathbb{E}[\mbox{log}_2 \Pi_{\widehat{\textbf{d}}}], 
\end{eqnarray}

\noindent where $\widehat{\textbf{d}}$ represents the quantized DCT coefficients, and $\Pi_{\widehat{\textbf{d}}}$ denotes the probability mass function. As shown in \cite{Balle17entropy}, $\widetilde{\textbf{d}} = \widehat{\textbf{d}} + \Delta \textbf{d}$ is a continuous relaxation of the probability mass function $\Pi_{\widehat{\textbf{d}}}$, where $\Delta \textbf{d}$ is additive i.i.d. uniform noise with the same minimum and maximum as the quantization bins. This means that the differential entropy of $\widetilde{\textbf{d}}$ can be used as an approximation of $\mathbb{E}[\log_2 \Pi_{\widehat{\textbf{d}}}]$. As suggested in \cite{balle2018variational}, the density function $\Pi_{\widetilde{\textbf{d}}}$ can be closely approximated by a non-parametric model consisting of a 3-layer neural network with 3 channels for each hidden layer followed by a Sigmoid non-linearity. Details of this approach are discussed in \cite{balle2018variational} and an online implementation is available in \cite{tfcompression}.

To imitate the actual JPEG encoder, we employ separate entropy estimators for each DCT channel (Luma and Chroma) and DC/AC coefficients (zero frequency and none-zero frequencies). This means that a total of four entropy estimators are trained in our framework. The overall entropy is the sum of these estimated entropies. We also use the DPCM (Differential Pulse Code Modulation) approach to encode the difference of adjacent DC components in JPEG blocks. Let $\textbf{d}^l_{k, i}$ be the $k$-th DCT component ($k\in \{0, \dots, 63\}$) of the $i$-th block ($i\in \{0, \dots, M\}$, where $M$ is the number of blocks) for the Luma ($l=0$) and Chroma ($l=1$) channels respectively, $E_{\theta^{l}_{*}}(\cdot)$ denote the approximate entropy parameterized by $\theta^{l}_{*}$, we have

\begin{equation}
 B^{l}_q(\textbf{x},\textbf{p}) = \sum_{i=0}^{M} [E_{\theta^{l}_{DC}}(\textbf{d}^l_{0, i}-\textbf{d}^l_{0, i-1}) + \sum_{k=1}^{63}E_{\theta^{l}_{AC}}(\textbf{d}^l_{k, i})],
 \label{eq:entropy2}
\end{equation}
where $\textbf{d}^l_{0, -1}:=0$, $B^{l}_q$, $l=\{0, 1\}$ are the estimated bit-rate for the Luma / Chroma channels respectively. The overall estimated bit-rate is given by $B_q = B^0_q + B^1_q$.

\begin{figure}[!ht]
\centering
        \includegraphics[ width=0.9\linewidth]{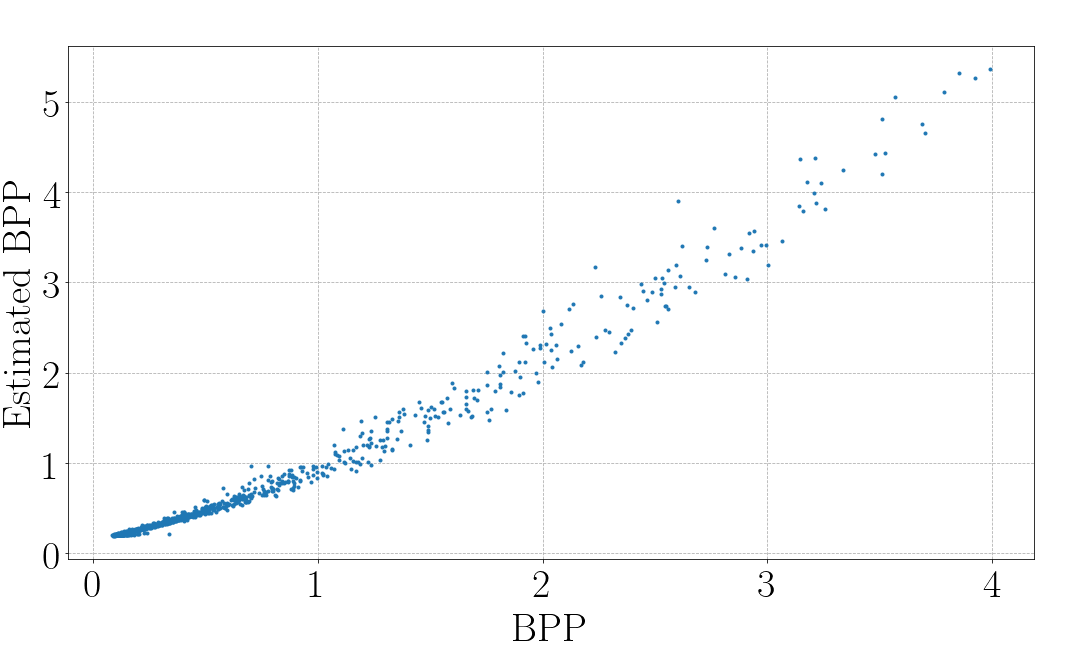} 
    \caption{Estimated BPP vs actual BPP generated from 100 sample images with varying quality factors. The Pearson correlation is $0.984$. }
    \vspace{-5 mm}
    \label{fig:entropy}
\end{figure}

Results from our entropy predictor and the actual JPEG bit-rates are presented in Figure~\ref{fig:entropy}. To generate these results, we randomly sampled 100 images from ImageNet test set and varied the JPEG quality factor in $[10,90]$ range. These results show a strong linear correlation of $0.984$ between the estimated and actual BPPs. We also observed that for BPP$\geq 2$, our method slightly over-estimates the actual bit-rate. 

    


\subsection{Classification Loss}

The function $A\left[\z,\y \right]: [\mathbb{R}^N \times \mathbb{R}^L]  \rightarrow \mathbb{R}^+$ denotes the standard softmax cross entropy for predictions from the image $\z$ with respect to the reference label $\y\in \mathbb{R}^L$. This function's evaluation includes within it the activation of the classification network 
in order to produce the predicted label, and the chosen loss between this label and the reference one.

\subsection{Overall Training Loss}

\noindent Using all the above differentiable ingredients, our loss function per image $\x$ and the quality factor $q$ is given by the following continuous function: 
\begin{eqnarray}\label{eq:formulation1}
LOSS_{image}(\p) & =&  c_r B_q(\x,\p ) + c_d \left\|C_q(\x,\p) - \x \right\|_2^2 + c_c A\left[C_q(\x,\p),\y\right], 
\end{eqnarray}
where $c_r,~c_d$ and $c_c$ are three weight coefficients governing the importance of the rate, distortion and classification losses, respectively. By modifying these weights we change the design goal of our optimization. Minimizing this function with respect to $\p$ via mini-batch gradient descent and back-propagation leads to the designed optimal quantization tables.

The above loss refers to the per-image case, where the quantization tables are best fitted for a single image $\x$ and a specific quality factor $q$. \blue{We should note that while this mode of operation makes perfect sense for the rate-distortion optimization, it is impractical when classification is involved (i.e. if $c_c>0$). This is due to the need to have the ground-truth label for the image in the optimization loss, information that is unavailable when a new image is given. Still, such a design goal for the quantization tables is of interest as it sets an upper-bound on the attainable accuracy when these tables are somehow image-adapted.}

When handling a corpus of $K$ images $\{\x_k\}_{k=1}^K$ and working in a range of quality factors $q\in \Omega$, the loss function will simply be a summation over these domains, 
\begin{eqnarray}\label{eq:formulation2}
LOSS_{universal}(\p) & =& \sum_{q\in \Omega} \sum_{k=1}^K \left\{ c_r B_q(\x_k,\p ) \right. \\
\nonumber & & \hspace{0.1in} \left. 
+ c_d \left\|C_q(\x_k,\p) - \x_k \right\|_2^2 
+ c_c A\left[C_q(\x_k,\p),\y_k\right] \right\}. 
\end{eqnarray}


\section{Results}
\label{Sec4:results}

In this section we discuss our experimental results. We optimize and evaluate our framework on the ImageNet benchmark~\cite{russakovsky2015imagenet}. In the following we first give a detailed overview of the experiment setup, and then go over the results for rate-distortion and rate-accuracy. We report our results from optimizing a single table-pair for all images (universal), and also the per-image case.  Ablation studies are also included to further elaborate on parameter choices made. The optimized quantization tables for each task are presented in the Appendix section.

\subsection{Experimental Setup}
For all experiments we use images of size $299\times299$ pixels (resized using bilinear interpolation) so as to be consistent with the input dimensions used for training the classification networks used hereafter. The quantization tables are optimized by minimizing the objective in Eq.~(\ref{eq:formulation2}). To evaluate the performance with the obtained tables, we generate the rate-distortion (or accuracy) curves by scaling the tables by a range of quality factors. The scaled tables are rounded and clipped to $\{1, \dots, 255\}$ before being used for compression. All images are compressed with \textsf{libjpeg-turbo}~\cite{libjpeg}. For each fixed quality factor, we average the bits-per-pixel (BPP), and the PSNR (or classification accuracy) to produce one point on the rate-distortion (or -accuracy) curves. 
For the universal case, the tables are optimized on the ImageNet training set, and evaluated on the eval set. For the per-image case, we randomly choose $1024$ images from the eval set and directly optimize on these. We note that the per-image case computes a single table for each image, and does not need to generalize to other images.

We tune the weights $c_r$ and $c_d$ (or $c_c$) in Eq.~(\ref{eq:formulation2}) so as to achieve the best overall performance across the full range of bit rates. The effect of adjusting these weights will be shown in the ablation study in Section~\ref{subsec:ablation}. For all experiments, we set the range of quality factors to be randomly sampled from $[10, 90]$ at optimization time. All models are optimized for 1M steps using ADAM with a learning rate of $10^{-4}$, and initialized with the default JPEG quantization tables. Also, our optimization batch size is set as $4$ across all experiments.



\subsection{Rate-Distortion Optimized Performance}
\label{subsec:rd}

\begin{figure}[]
    \centering
    \begin{tabular}{c c}
        \includegraphics[ width=0.50\linewidth]{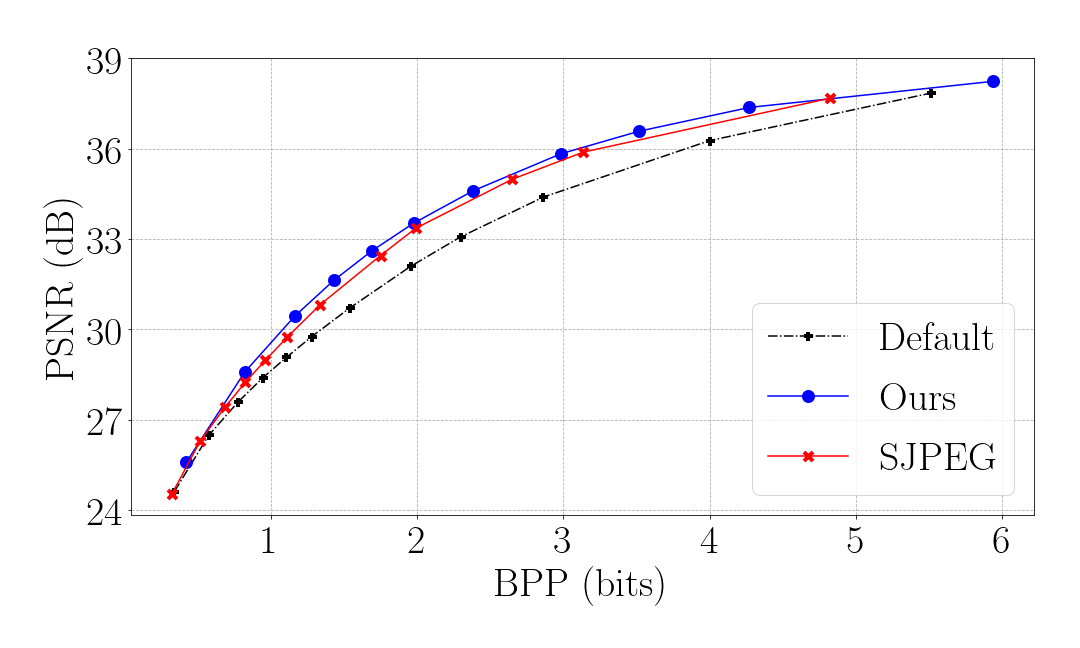}& \hspace{-0.15in} \includegraphics[ width=0.50\linewidth]{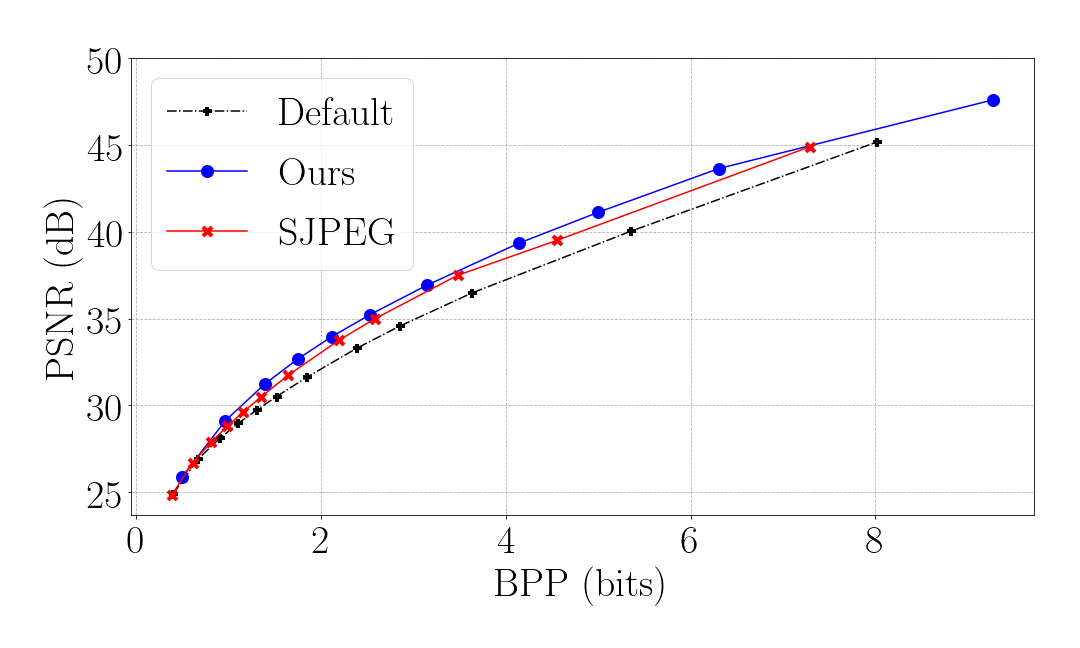} \\
    \end{tabular}
    \caption{PSNR vs. Bits-Per-Pixel (BPP) for the universally optimized quantization tables compared to the default tables evaluated on ImageNet. Left: YUV 420. Right: YUV 444. }
    \vspace{-6 mm}
    \label{fig:rate_distortion_universal}
\end{figure}

This start by presenting the results for the rate-distortion optimization task. For this experiment the loss weights in Eq.~(\ref{eq:formulation2}) are set as $c_r=c_d=1$ and $c_c=0$. The optimized rate-distortion curves are shown in Figure~\ref{fig:rate_distortion_universal}, considering both YUV420 and no chroma sub-sampling (YUV444). As can be seen, compression with the proposed quantization tables leads to a consistent improvement in PSNR for a fixed bit-rate. This improvement is more significant for $\mbox{BPP}\in[0.5,3.5]$. For instance, at $2$ bits-per-pixel, our customized tables outperform the default ones by nearly $1.5$dB. Conversely, our tables achieve a $25\%$ reduction in file size while maintaining the same image quality at PSNR=$36$dB. Samples of compressed images using the optimized quantization tables and the default ones are shown in Figure~\ref{fig:rd_improvements}. As can be seen, the found quantization tables reduce the file-size of both images while preserving the image quality, assessed via PSNR. All images are compressed with the libjpegturbo library\cite{libjpeg}.

\subsubsection{Comparison With SJPEG}

\begin{figure}[]
    \centering

    \begin{tabular}{c c c c}
    Default & Ours  & Default & Ours \\
    \includegraphics[ width=0.20\linewidth]{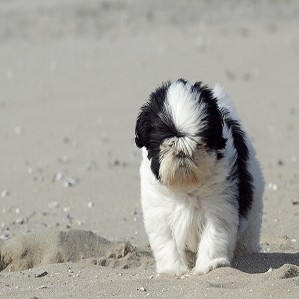} &
    \includegraphics[ width=0.20\linewidth]{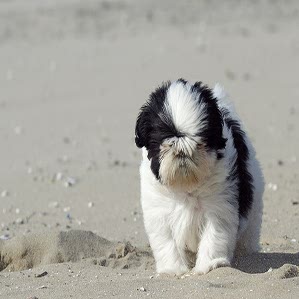}  & 
    \includegraphics[ width=0.20\linewidth]{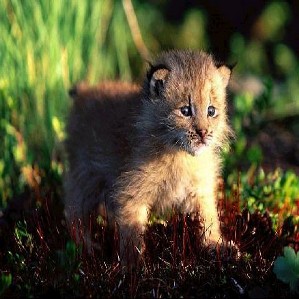} &
    \includegraphics[ width=0.20\linewidth]{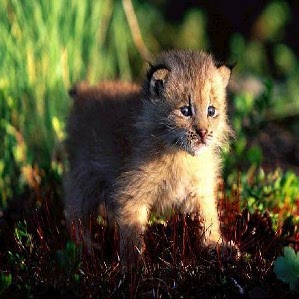}  \\
    PSNR=$39.22$dB  & PSNR=$39.48$dB   &   PSNR=$37.07$dB & PSNR=$37.26$dB         \\
    BPP=$1.90$ & BPP=$1.44$ ($-24.2\%$) &  BPP=$2.88$ & BPP=$2.31$ ($-19.5\%$)     \\

    \includegraphics[ width=0.20\linewidth]{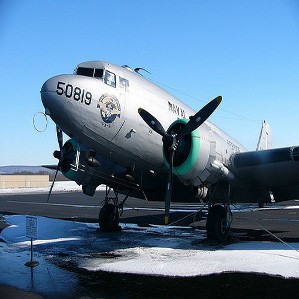} &  \includegraphics[ width=0.20\linewidth]{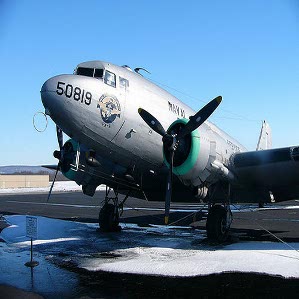}  &  \includegraphics[ width=0.20\linewidth]{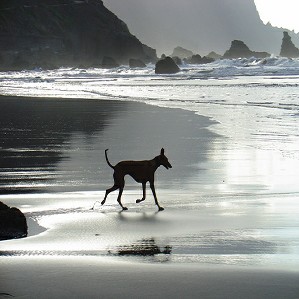} & 
    \includegraphics[ width=0.20\linewidth]{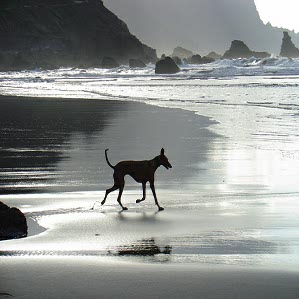}  \\   
    PSNR=$35.81$dB & PSNR=$35.82$dB   & PSNR=$35.77$dB & PSNR=$35.81$dB            \\ 
    BPP=$2.45$ & BPP=$1.87$ ($-23.6\%$)  & BPP=$2.41$    & BPP=$1.87$ ($-22.1\%$) 
    \end{tabular}
    
    \caption{Examples showing the effect obtained by our optimized quantization tables versus the default ones for the rate-distortion optimization. In all four examples the PSNR remains almost unchanged, while the rates show a significant savings. All cases refer to $q=90$ (quality factor) used in the default case.}
    \label{fig:rd_improvements}
\end{figure}

We compare our method with SJPEG~\cite{sjpeg}, an open-source JPEG compression library that supports image-adaptive optimizations of JPEG quantization tables. On a high-level, SJPEG performs coordinate descent to minimize the rate-distortion objective $D + \lambda R$, similar to \cite{fung1995design}, where $\lambda$ is dynamically chosen to approximate the slope of the current point on the R-D curve.
 Figure~\ref{fig:rd_improvements} shows our universal method outperforms SJPEG, especially in the range of BPP$=[2.0, 4.0]$. We emphasize that SJPEG~\cite{sjpeg} adapts the quantization table \emph{per image and per quality value}, whereas our method uses a single table for all quality values and images. For a fair comparison, we only use SJPEG to optimize the quantization table, and apply libjpegturbo with the optimized tables from SJPEG for the final stats.
 
\subsubsection{Per-Image Optimized Tables }
Next, we compare the performance of the universally optimized tables with the per-image obtained ones. The hyperparameters $c_r$ and $c_d$ are set exactly as in the universal case and the optimization uses the same parameters. The results are evaluated on the same set of $1024$ randomly sampled images from the ImageNet validation set, which were used also in the universal and the default tests, and the shown rate-distortion curve is an average over this image set.



\begin{figure}[]
    \centering
    \begin{tabular}{c c} \includegraphics[ width=0.50\linewidth]{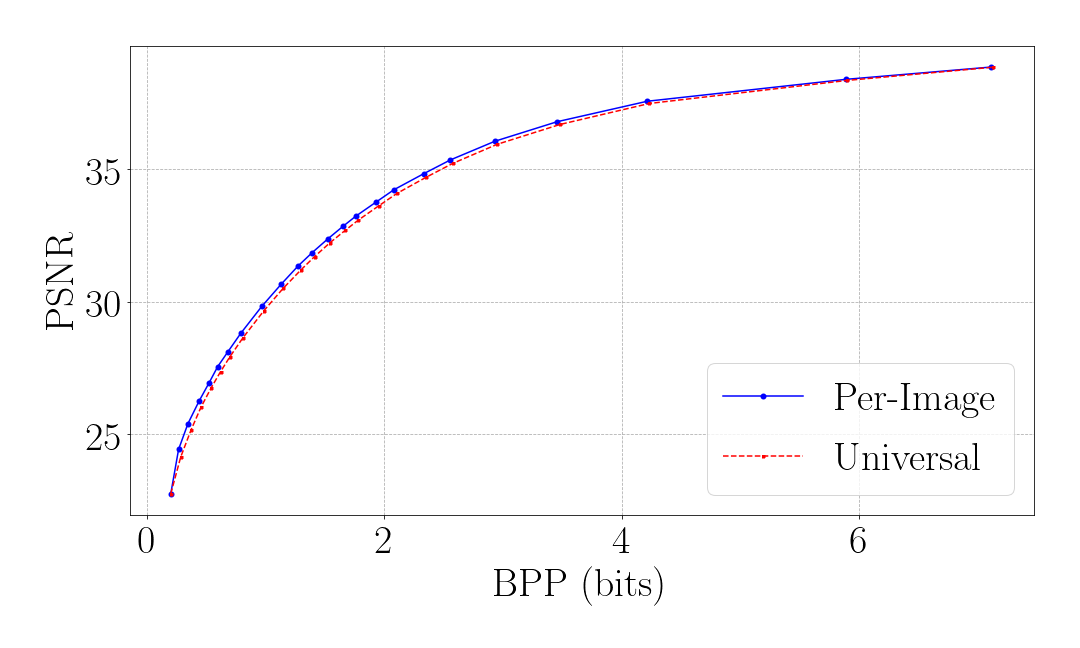} & \hspace{-0.15in}
    \includegraphics[ width=0.50\linewidth]{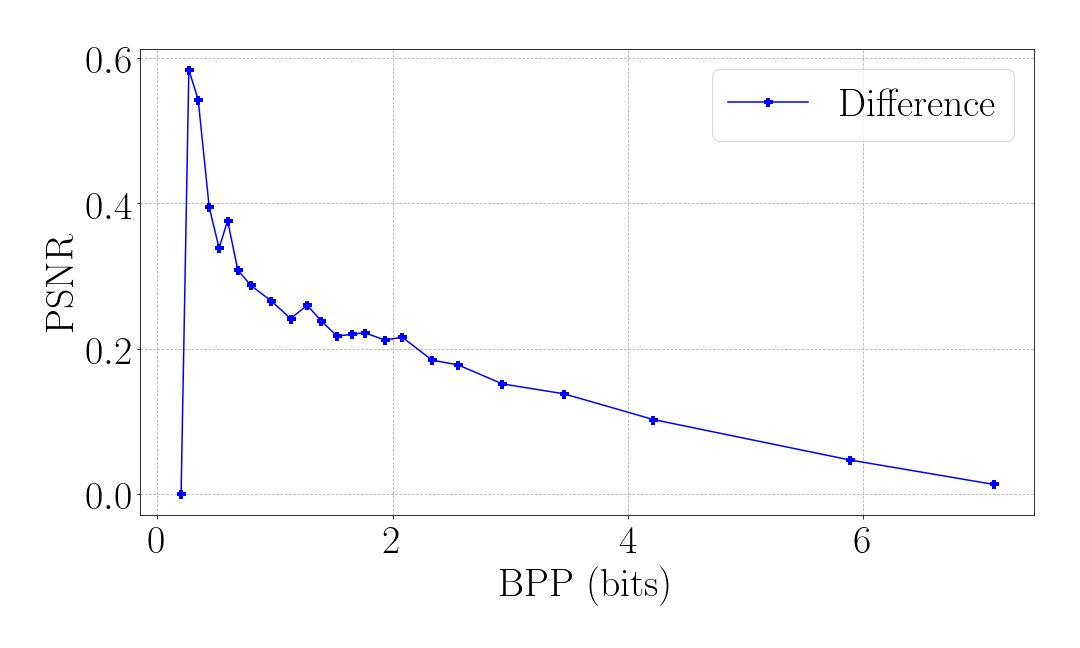} \\
    (a) & (b)
    \end{tabular}
    \caption{(a) Rate-distortion curves for the default, the universally optimized and the per-image optimized tables, all evaluated on a randomly sampled subset of $1024$ images from the ImageNet eval set. b) A comparison of our universal tables with SJPEG, both evaluated on the ImageNet eval set. These graphs correspond to YUV420 chroma sub-sampling.}
    \label{fig:per_image_and_sjpeg}
\end{figure}

As shown in Figure~\ref{fig:per_image_and_sjpeg}, optimizing the quantization tables on a per-image basis leads to a further improvement of around $0.6$dB compared to the universal case for lower bit rates, and around $0.2$dB for higher bit rates.

\subsection{Classification Optimized Performance}

In the context of rate-accuracy (R-C) performance, we report the results of our method on three networks:  ResNet-V2-101~\cite{he2016identity},  ResNet-V2-50~\cite{he2016identity}, and MobileNetV3~\cite{howard2019searching}. To optimize the JPEG quantization tables for a classification task, we set the loss weights in Eq.~(\ref{eq:formulation2}) to $c_d=0$, $c_c=1.0$, and set $c_r=10.0$. All tables are optimized on the ImageNet training set, and tested on its evaluation set. 

\begin{figure}[!ht]
    \centering
    \begin{tabular}{c c} \includegraphics[ width=0.50\linewidth]{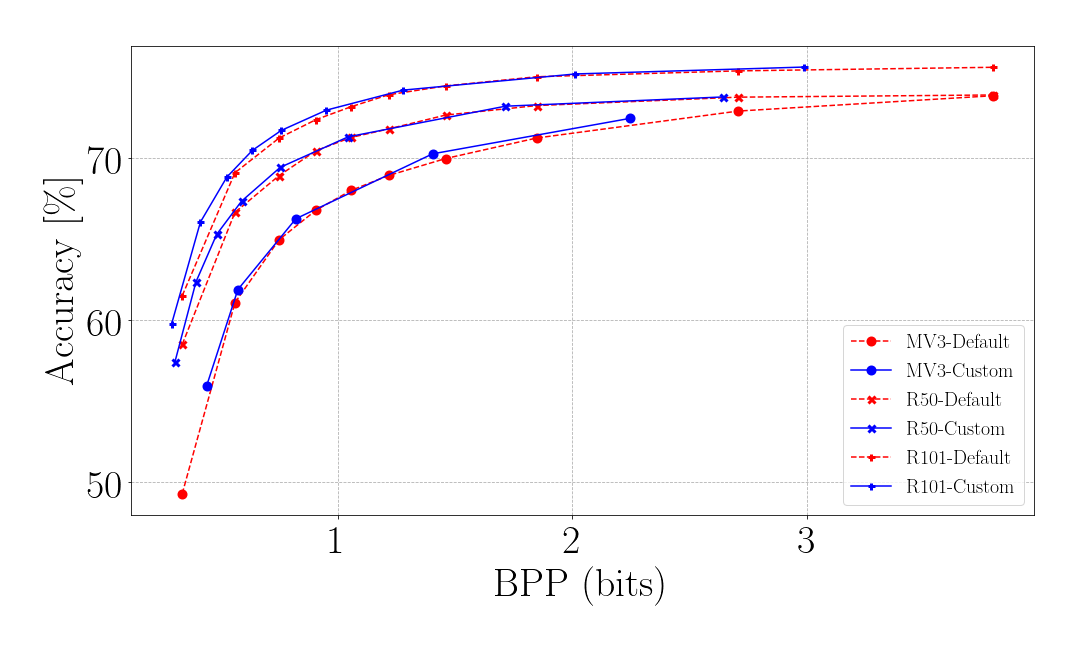} & \hspace{-0.15in}
    \includegraphics[ width=0.50\linewidth]{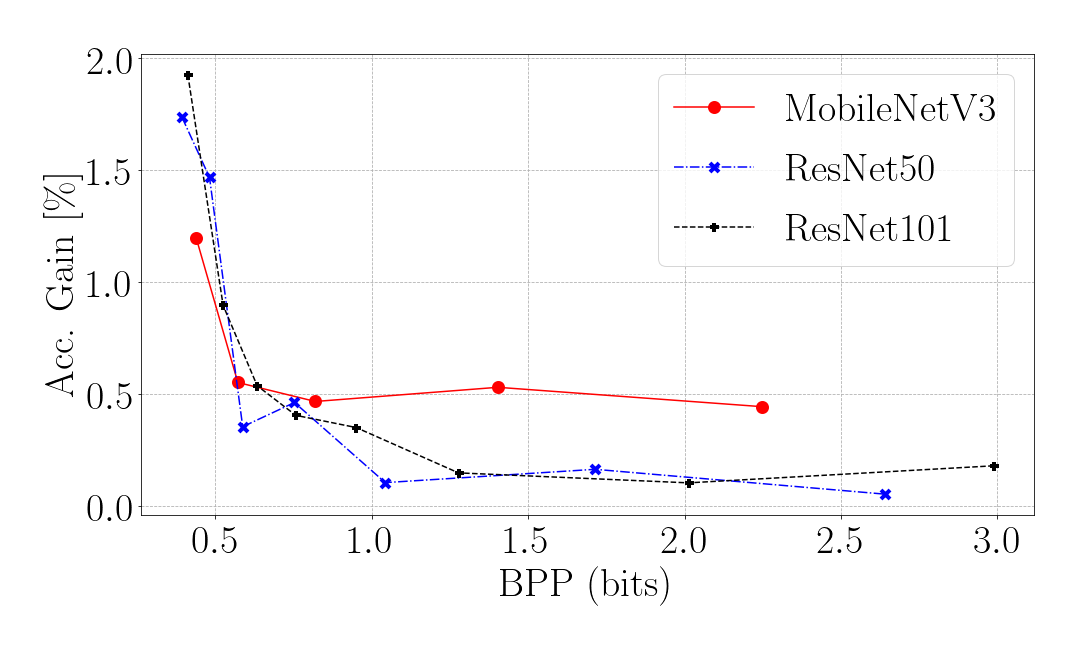} \\
    (a) BPP vs Accuracy for default table & (b) Accuracy Gain 
    \end{tabular}
    \caption{Classification Accuracy gain versus Bits-per-pixel (BPP) for YUV420 chroma sub-sampling. Left: Accuracy vs BPP for the default quantization table. Right: Accuracy gain of the universal table across a range of BPPs. }
    \label{fig:global_bpp_acc_420}
\end{figure}

Figure~\ref{fig:global_bpp_acc_420} presents the rate-accuracy curves for the default and universal tables. The attained gains in accuracy is also plotted for easier visualization. We note that the gain is larger for MobileNetV3,  likely due to the fact that it has a smaller network capacity compared to ResNet, and hence is more sensitive to compression artifacts.

\begin{figure}[!ht]
    \centering
    \begin{tabular}{c c c c}
      Default & Ours & Default & Ours\\
         \includegraphics[ width=0.20\linewidth]{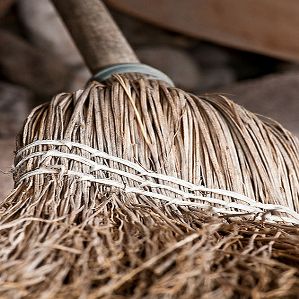}      &
         \includegraphics[ width=0.20\linewidth]{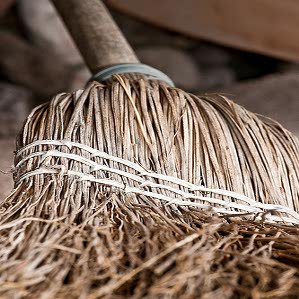}    &          \includegraphics[ width=0.20\linewidth]{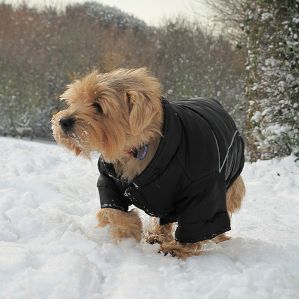}      &
         \includegraphics[ width=0.20\linewidth]{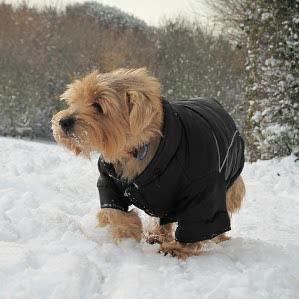}     \\    
         ``swab" & ``broom" & ``Dinmont" & ``Norfolk terrier"   \\       
         BPP = $2.616$ & BPP = $2.569$ &  BPP =1.649 & BPP =1.683\\
         Prob = $0.275$ & Prob = $0.301$ & Prob = $0.598$ & Prob = $0.416$            
    \end{tabular}
    \caption{Examples of wrong classification corrected by our optimized tables. $Q$ is set to $50$ for the default table and tuned for the custom table such that the file size is slightly smaller. The predicted label, BPP, as well as the probability scores are shown below.}
    \label{fig:classification_improvement}
\end{figure}

Figure~\ref{fig:classification_improvement} brings three examples where our optimized tables fix an incorrect classification, while using a lower bit-rate. The probabilities reported are obtained from the classification network, representing the confidence in the decisions made.
Though visually similar, we can see (best viewed zoomed in) that images from the obtained tables enhance certain texture regions of the image. 

\subsubsection{Comparison with Sorted Random Search (SRS)}
\vspace{-5pt}
Sorted Random Search~\cite{li2020optimizing} is a simple yet effective method for optimizing the quantization tables for classification. For a fair comparison, we randomly search for 1000 tables as in~\cite{li2020optimizing}, and randomly sample 5000 images from the training set for each candidate's rate-accuracy evaluation. This ensures that SRS has seen roughly the same number of examples as our mini-batch optimization. We choose a Pareto optimal table closest to 0.8 BPP for MobileNetV3.
Figure~\ref{fig:per_image_bpp_acc_420} (b) shows that our table outperforms SRS on a range of BPP values. A possible explanation is that the SRS method needs a sufficiently large and well-sampled (such as the MatchedFrequency strategy used in~\cite{li2020optimizing}) set per-epoch, whereas our method fully takes advantange of the benefits of mini-batch back-propagation and results in better performance.


\subsubsection{Per-Image Results}

\begin{figure}[]
    \centering
    \begin{tabular}{c c}
        \includegraphics[ width=0.50\linewidth]{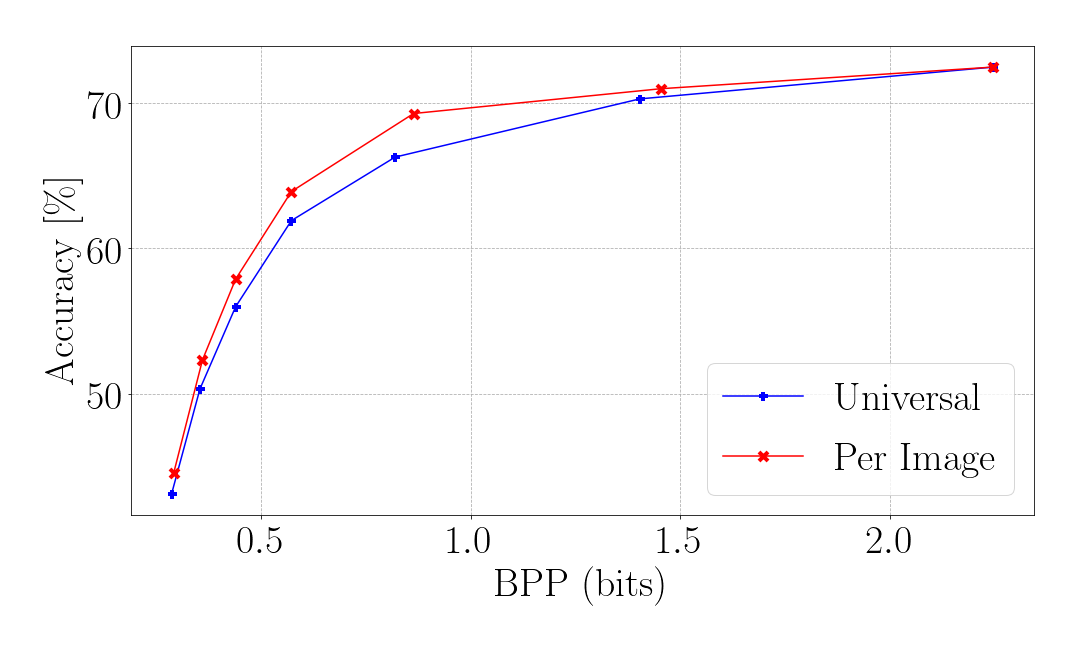}&  \hspace{-0.15in}
        \includegraphics[ width=0.50\linewidth]{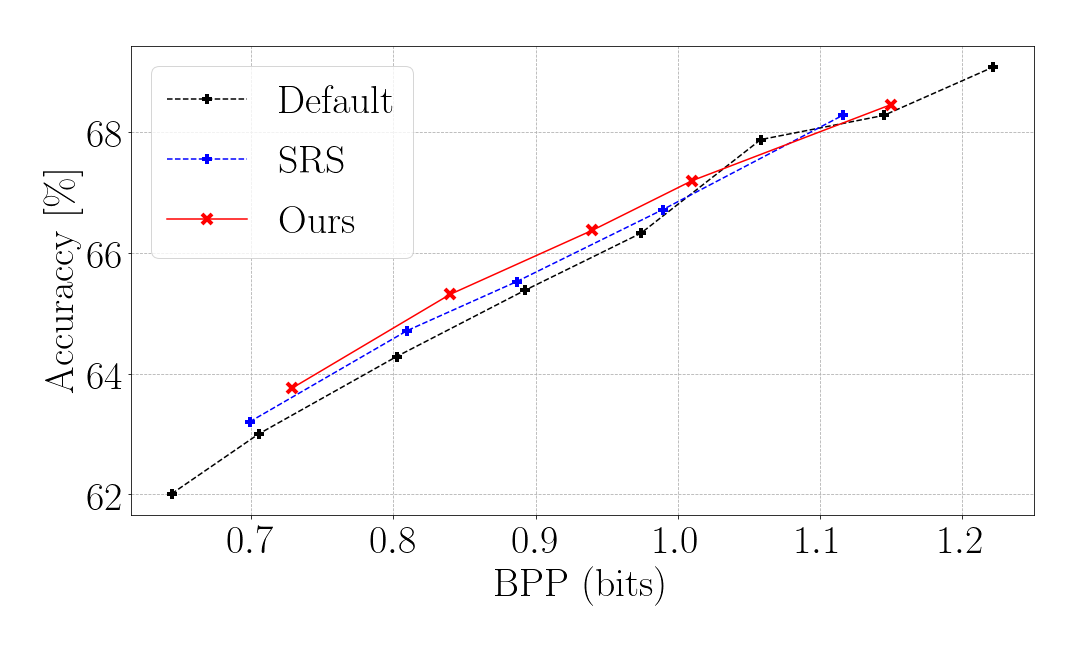} \\
        (a) & (b)
    \end{tabular}
    \caption{Left: Bits-per-pixel vs classification accuracy for per-image and universal tables. Reported results are on 1024 randomly sampled images from the ImageNet evalset, with chroma sub-sampling. Right: Accuracy of the default table, SRS, and our method on 10k randomly sampled images from the ImageNet eval set.}
    \label{fig:per_image_bpp_acc_420}
    \vspace{-10pt}
\end{figure}

We turn to present the per-image results. The same subset of 1024 images as in Section~\ref{subsec:rd} is used for evaluation. Figure~\ref{fig:per_image_bpp_acc_420} shows a large increase in terms of accuracy. As already mentioned in Section~\ref{Sec3:Proposed}, the recognition loss for the per-image tables contains the ground truth label of the image, and hence the per-image method is not a practical approach \blue{(unless the ground truth label is available, such as in identification tests)}. Nevertheless, the result still has value as it shows the performance limit of adapting the JPEG quantization table for the purpose of boosting the classification accuracy.

\begin{figure}[!ht]
    \centering
    \begin{tabular}{c c}
        \includegraphics[ width=0.50\linewidth]{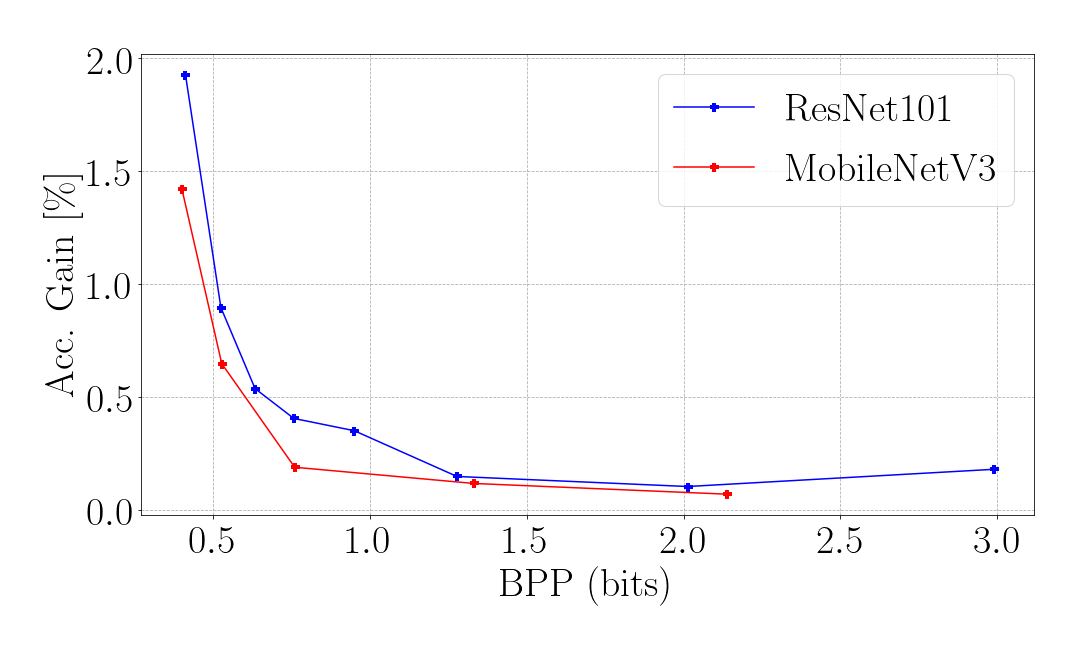}&  \hspace{-0.15in}
        \includegraphics[ width=0.50\linewidth]{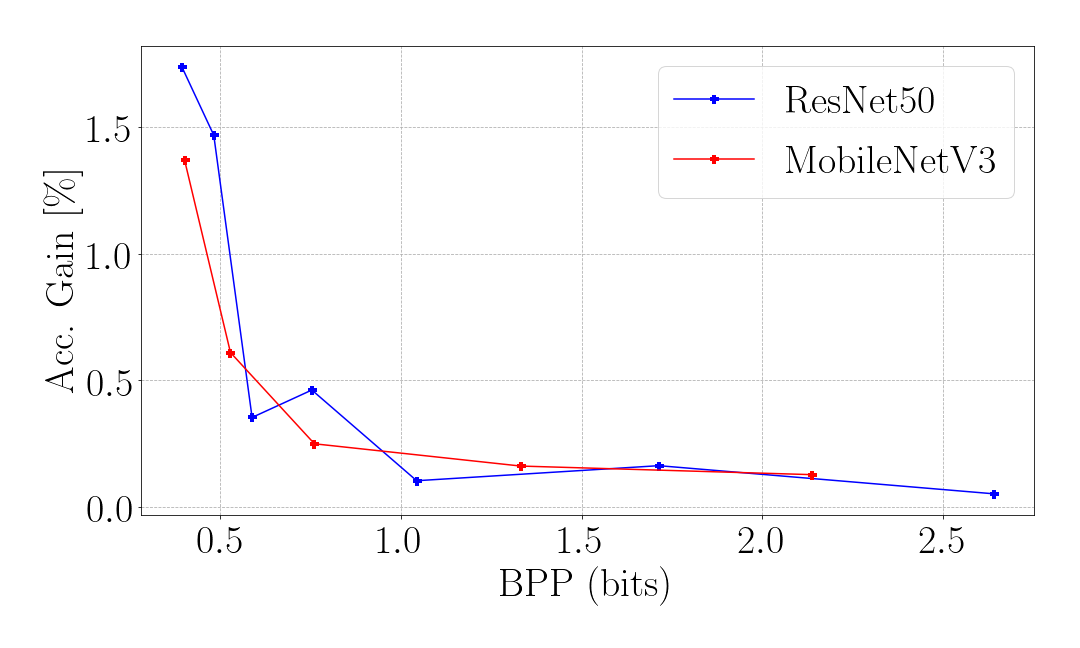} \\
    \end{tabular}
    \caption{Generalization of optimized tables to other networks: These graphs show the accuracy gain (over the default tables) with ResNet101 and ResNet50 while using tables optimized for MobileNetV3. All results are evaluated on the full ImageNet eval set.}
    \label{fig:generalize}
\end{figure}
\subsubsection{Generalization to Other Networks}
Next, we show that the accuracy performance generalizes across different networks. Figure~\ref{fig:generalize} illustrates the classification accuracy against BPP evaluated using ResNet101, for a quantization tables optimized on MobileNetV3. Observe that the classification accuracy improves across a wide range of rates, even though the table was computed with a different network loss.

\subsection{Ablation Study}
\label{subsec:ablation}

\begin{figure}[]
    \centering
    \begin{tabular}{c c} \includegraphics[ width=0.50\linewidth]{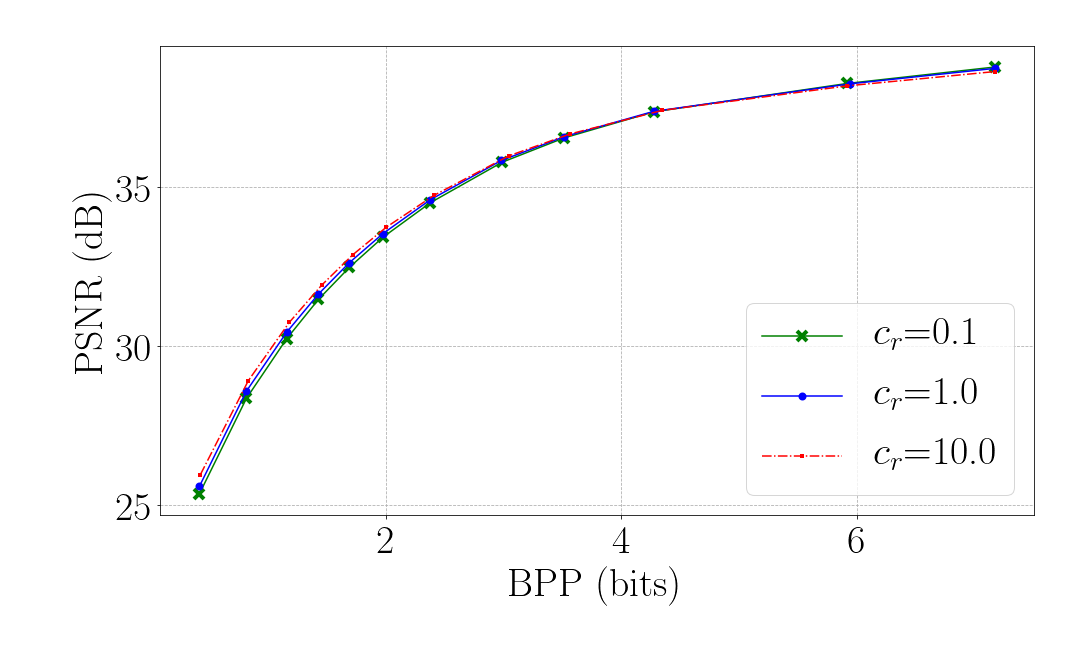} & \hspace{-0.15in}
    \includegraphics[ width=0.50\linewidth]{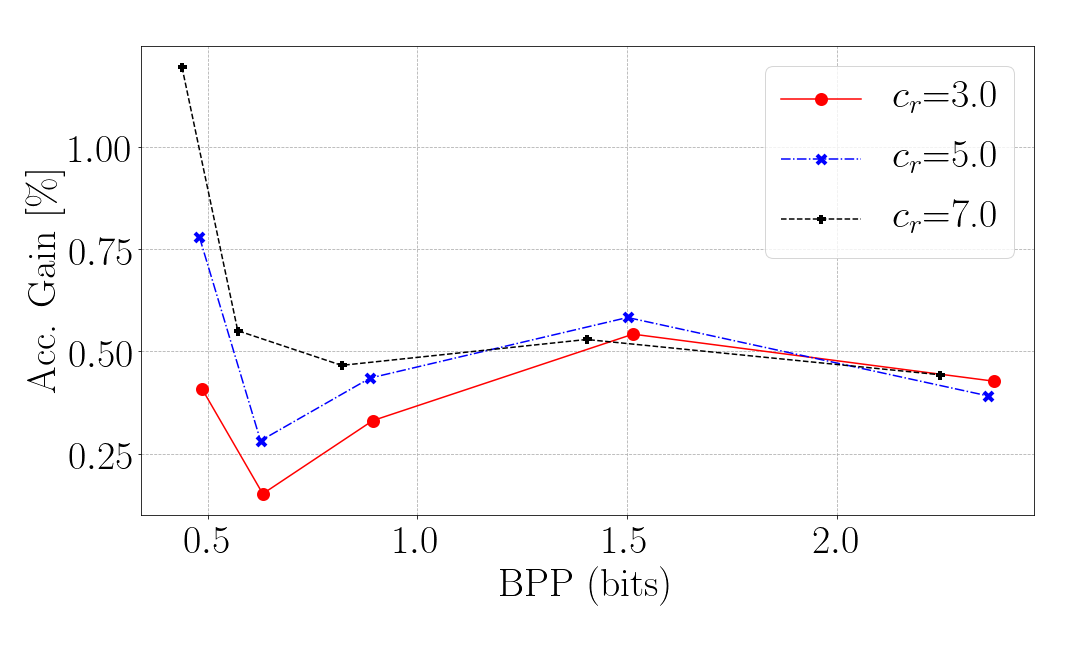} \\
    (a) Rate-Distortion & (b) Rate-Accuracy (MobileNetV3) \\
    \includegraphics[width=0.50\linewidth]{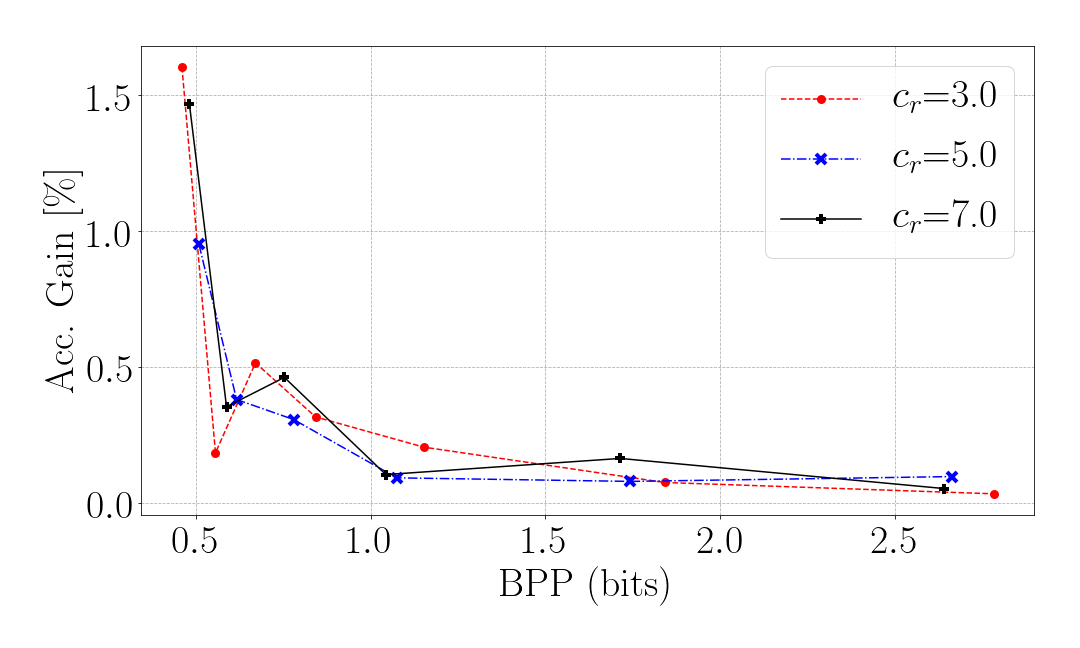}& 
     \hspace{-0.15in}
     \includegraphics[ width=0.50\linewidth]{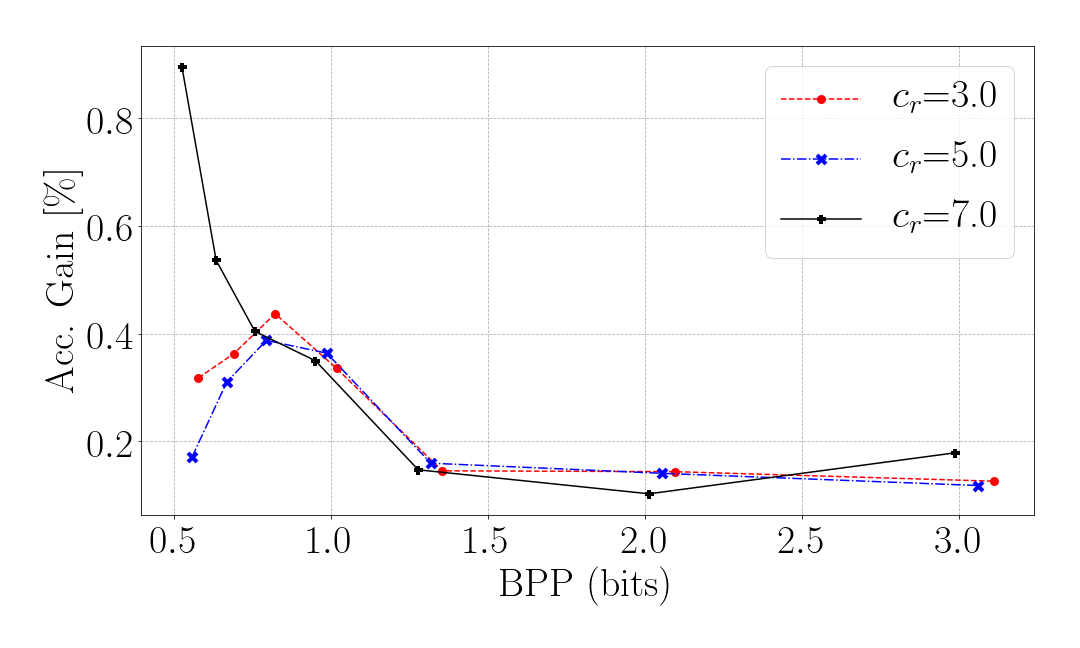} \\
    (c) Rate-Accuracy (ResNet50) & (d) Rate-Accuracy (ResNet101)
    \end{tabular}
    \caption{Varying the rate weight $c_r$ in the loss-function for rate-distortion optimization and classification tasks.}
    \label{fig:hyperparams_classification}
\end{figure}

Figure~\ref{fig:hyperparams_classification} (a) shows the effect of adjusting the hyperparameter $c_r$ relative to $c_d$ for rate-distortion optimization. We vary $c_r$ from $0.1$ to $10.0$, and fix $c_d=1.0$. We see that for larger values of $c_r$, the quantization table in general performs better for lower bit-rates, and worse for higher ones. We also observe that the performance is not particularly sensitive to these hyperparameters, as indicated by the proximity of the R-D curves for $c_r=1.0$ and $c_r=10.0$. All experiments are optimized and evaluated under the same setup (except $c_r$) as in Section~\ref{subsec:rd}, with Chroma sub-sampling.

In Figure~\ref{fig:hyperparams_classification}  (b)-(d), a similar analysis on the loss weights is provided by varying $c_r$ and fixing $c_c=1.0$, while considering three classification networks -- MobileNetV3, ResNet50 and ResNet101. Referring to MobileNetV3, we see that larger values of $c_r$ result in a higher gain for lower BPP, and less so when the BPP is high. 

\section{Conclusion}
Lossy image compression methods trade file-size (rate) for faithfulness to the original image content (distortion). A third performance measure influenced by such coding is classification accuracy, as affected by the induced error in the image. In this paper we offer an investigation of the tradeoff between these three performance measures, analyzed in the context of the JPEG compression algorithm. We show that JPEG's quantization tables can be optimized for better rate-distortion or rate-accuracy performance. Our work introduces two modes of optimization -- universal and per-image. The universal mode of work targets a single set of tables for all training images, so as to replace the default ones. The per-image model assigns the tables for each image, further boosting the rate-distortion behavior. In our future work we aim to train a deep neural network that produces the best quantization tables for any incoming image, this way producing a highly effective per-image treatment. In the context of rate-accuracy behavior, the per-image mode we present here will stand as an upper-bound to the achievable results. Our future plans also include a similar treatment of other compression standards.


\section{Appendix}

Table~\ref{tab:quant_rd} provide the universally optimized quantization tables for rate-distortion. The optimized tables are computed with chroma sub-sampling, using the hyperparameters reported in Section~\ref{Sec4:results}. The values of the universally optimized are reported in floats. To produce the rate-distortion (accuracy) curves Section~\ref{Sec4:results}, the tables are first scaled, and then rounded and clipped to $\{1, \dots, 255\}$ before passing to the JPEG encoder.

In Table~\ref{tab:quant_rd}, the learned tables for rate-distortion quantize the chroma channels much less compared to the default tables. We note that this is due to the default tables being designed for the Human Visual System (HVS) instead of image distortion, since HVS is less sensitive to the chroma channel. In fact, our framework can easily adaptive to optimizing for visual perception by adding perceptual losses~\cite{johnson2016perceptual}, or by manually weighting the luma and chroma bit-rate losses in Equation~\ref{eq:entropy2}. Experimentation for perceptual metrics is outside the scope of this paper, and could be an interesting direction for future work. 


\begin{table}[!ht]
    \begin{minipage}{.5\linewidth}
      \centering \scriptsize{
    \begin{tabular}{|r |r |r |r |r |r |r |r|} \hline
        16.0& 14.9& 14.2& 14.8& 15.6& 17.7& 18.9& 20.0 \\ \hline
        15.1& 14.5& 14.5& 14.9& 15.6& 19.8& 19.7& 18.7 \\ \hline
        14.8& 14.3& 14.5& 15.4& 17.3& 19.3& 20.7& 18.6 \\ \hline
        14.4& 14.6& 15.1& 15.8& 18.5& 23.2& 21.9& 19.2 \\ \hline
        14.6& 14.9& 16.8& 19.1& 20.4& 26.1& 24.9& 21.0 \\ \hline
        15.1& 16.3& 18.8& 19.8& 21.9& 25.0& 26.1& 22.9 \\ \hline
        18.3& 19.9& 21.6& 22.6& 24.7& 27.2& 26.8& 23.9 \\ \hline
        21.3& 23.6& 23.8& 23.9& 25.8& 23.7& 24.0& 23.3 \\ \hline
    \end{tabular}}
    \end{minipage}%
    \begin{minipage}{.5\linewidth}
      \centering \scriptsize{
    \begin{tabular}{|r |r |r |r |r |r |r |r|} \hline
        14.3& 14.9& 14.7& 16.9& 23.5& 22.8& 22.3& 21.9\\ \hline
        14.9& 14.1& 13.9& 18.6& 22.7& 22.0& 21.6& 21.3\\ \hline
        14.6& 13.9& 17.2& 22.8& 22.2& 21.6& 21.2& 21.0\\ \hline
        16.8& 18.5& 22.8& 22.3& 21.7& 21.2& 20.8& 20.6\\ \hline
        23.4& 22.6& 22.1& 21.7& 21.2& 20.7& 20.4& 20.3\\ \hline
        22.6& 21.9& 21.5& 21.1& 20.7& 20.3& 20.0& 19.9\\ \hline
        22.1& 21.4& 21.1& 20.7& 20.3& 20.0& 19.8& 19.7\\ \hline
        21.8& 21.1& 20.8& 20.5& 20.2& 19.9& 19.7& 19.6\\ \hline
    \end{tabular}}
    \end{minipage} 
    \caption{Universally optimized tables for rate-distortion performance. Left: Luma, Right: Chroma.}    
    \label{tab:quant_rd}
\end{table}


\bibliographystyle{splncs}
\bibliography{references}

\end{document}